# The Neptune Trojans – a new source for the Centaurs?

Jonathan Horner[1] and Patryk Sofia Lykawka[2*]
[1] Dept. of Physics and Astronomy, The Open University, Walton Hall, Milton Keynes, MK7 6AA, UK; e-mail: j.a.horner@open.ac.uk
[2] International Center for Human Sciences (Planetary Sciences), Kinki University, 3-4-1 Kowakae, Higashiosaka, Osaka, 577-8502, Japan

---

[*] Previous address: Kobe University, Dept. of Earth and Planetary Sciences, Kobe, Japan

## ABSTRACT

The fact that the Centaurs are the primary source of the Short Period Comets is well established. However, the origin of the Centaurs themselves is still under some debate, with a variety of different source reservoirs being proposed in the last decade. In this work, we suggest that the Neptune Trojans (together with the Jovian Trojans) could represent an additional significant source of Centaurs. Using dynamical simulations of the first Neptune Trojan discovered (2001 QR322), together with integrations following the evolution of clouds of theoretical Neptune Trojans obtained during simulations of planetary migration, we show that the Neptune Trojan population contains a great number of objects which are unstable on both Myr and Gyr timescales. Using individual examples, we show how objects that leave the Neptunian Trojan cloud evolve onto orbits indistinguishable from those of the known Centaurs, before providing a range of estimates of the flux from this region to the Centaur population. With only moderate assumptions, it is shown that the Trojans can contribute a significant proportion of the Centaur population, and may even be the dominant source reservoir. This result is supported by past work on the colours of the Trojans and the Centaurs, but it will take future observations to determine the full scale of the contribution of the escaped Trojans to the Centaur population.

# 1 INTRODUCTION

It is well established that the Centaurs, small bodies on orbits which bring them closer to the Sun than Neptune, but which stay beyond the orbit of Jupiter, are the primary source of short-period comets (Duncan, Quinn & Tremaine 1988; Horner, Evans & Bailey 2004b). However, the origin of the Centaurs themselves is still under debate. The initial reasoning behind the suggestion that a belt of objects lay beyond the orbit of Neptune (Edgeworth 1949; Kuiper 1950) was that there needed to be an asteroid belt like collection of cometary nuclei, held in cold storage beyond the orbit of that giant planet, which sourced fresh material to replace the comets lost over time. This idea, that objects from the trans-Neptunian region are the primary source of Centaurs, still holds sway today, with various authors suggesting links between the Scattered Disk, the Edgeworth-Kuiper (EK) belt, and even the inner Oort cloud and the Centaurs[1] (Holman & Wisdom 1993; Duncan & Levison 1997; Emel'yanenko, Asher & Bailey 2005). Whatever the source region of Centaurs may be, these objects have been shown to evolve on highly unstable and chaotic orbits, with typically mean half-lives ranging from a few hundred thousand years (for the least stable) to tens of Myr (for the most stable) (Tiscareno & Malhotra 2003; Horner, Evans & Bailey 2004a, 2004b; di Sisto & Brunini 2007).

In this work, we argue that the newly discovered Neptune Trojans could contribute a significant number of objects to the Centaur population over the age of the Solar system. Based on simulations of the observed Neptune Trojans, and of swarms of theoretical objects resulting from simulations of planetary migration, both of which show that the Neptune Trojan region population must contain large numbers of objects that are dynamically unstable over the lifetime of the Solar system, we illustrate how such objects can evolve from orbits initially bound within the Neptunian 1:1 mean-motion resonance (MMR) to ones similar to those observed in the current Centaur population. We use a number of examples of the long-term orbital evolution of Neptune Trojan objects initially located on orbits within the observational-error ellipse of 2001 QR322, the first Neptune Trojan discovered (Chiang et al. 2003), to demonstrate that such transfer from the Neptune Trojan cloud to Centaur orbits is perfectly compatible with the currently observed objects. It should be noted that 2001 QR322 has a long-arc orbit (1450 days, taken from the AstDys website in January 2009), and as such has particularly small orbital uncertainties, and so such evolution is not merely the result of a poorly-determined orbit for that presumably primordial object (Brasser et al. 2004).

It is currently thought that the Neptune Trojan population is at least as numerous as the main belt asteroids, and that the Trojans may outnumber the asteroids by an order of magnitude, or more (Sheppard & Trujillo 2006). As such, it is likely that any dynamically unstable component to Neptune's Trojan population would initially have numbered at least as many objects as their stable counterparts, and that the ongoing decay of these objects would continually feed fresh material into the *cis*-Neptunian region to become Centaurs and eventually short-period comets.

In section 2, we briefly describe the modelling carried out of the orbits of Neptune Trojans, before detailing our results in section 3, and concluding in section 4 with a discussion of the implications of our work, and avenues for further exploration.

# 2 MODELLING

We have carried out detailed dynamical studies of the formation and evolution of Neptune Trojans (Lykawka et al. 2009a, Lykawka et al. 2009b, Lykawka & Horner 2009). That work considers firstly the evolution of the Neptune Trojan population as the giant planet migrates outward through a planetesimal disk to its current location (Lykawka et al. 2009a), following both the capture of fresh material and the transport of pre-formed Trojans. In that work, four distinct dynamical scenarios were considered. In each case, the Solar system started the integrations in a significantly compacted form, and the giant planets migrated to their current locations in a smooth manner, according to

$$a_k(t) = a_k(F) - \delta a_k \exp(-t/\tau) \qquad (1)$$

where $a_k(t)$ is the semi-major axis of the planet after time $t$, $a_k(F)$ is the final (current) value of the semi-major axis, and $\tau$ is a constant determining the rate of migration of the planet. We examined both rapid and slow migration ($\tau = 1$, 10 Myr, with the planets taking a total of $5\tau$ years to reach their current heliocentric distances) for scenarios in which the initial semi-major axis of Neptune was set at 18.1 and 23.1 AU. Jupiter and Saturn began each of the four scenarios at heliocentric distances of 5.4 and 8.6 AU, respectively, while Uranus started at a variety of locations (between 12.2 and 14.7 AU when Neptune was placed at 18.1 AU, and between 14.8 and 16.6 AU when Neptune began at 23.1 AU).

---

[1] Trans-Neptunian objects (TNOs) in the Edgeworth-Kuiper belt, located beyond Neptune, exhibit orbits with semi-major axes, *a*, less than ~50 AU, and perihelia beyond ~35 AU, whereas TNOs in the Scattered Disk typically have semi-major axes greater than 50 AU, and/or perihelia closer than ~35 AU (but still beyond that of Neptune). Inner Oort cloud members are typically categorised as objects with aphelia between 1,000 and 10,000 AU (Morbidelli & Brown 2004; Brasser, Duncan & Levison 2006 and references therein; Lykawka & Mukai 2007a).

In each test, we followed the evolution of both pre-formed Neptunian Trojans (objects on dynamically cold orbits, trapped in Neptune's 1:1 MMR at the start of the integrations) and a vast swathe of objects located in a trans-Neptunian disk, tracking each individual test particle until it was either ejected from the system or collided with one of the massive bodies. For more detail on the integrations carried out, we direct the interested reader to Lykawka et al. (2009a).

We found that that smooth planetary migration, considered over a range of heliocentric distances, produces extended clouds of Neptune Trojans. As a follow up to that work, we took the test particles that made up the extended Trojan clouds at the end of the integrations, and used them as the seeds of a far larger Trojan population. The long term evolution of the bodies in that population was followed under the influence of the planets Jupiter, Saturn, Uranus and Neptune for a period of 1 Gyr, which allowed the determination of both the overall dynamical stability of the post-migration clouds, and the fates of any objects which left them. Again, the test particles involved were followed until they either collided with a massive body or were ejected from the Solar system. That work showed that the great majority of the objects in the extended Trojan clouds lay on orbits that are dynamically unstable on timescales of several tens to hundreds of Myr.

Previous studies also support the existence of a non-negligible smaller fraction of Trojans leaving the Trojan cloud on Gyr timescales (e.g., Nesvorny & Dones 2002). This instability yields an ongoing flux of material out of the Trojan clouds onto Neptune-encountering orbits – in other words, it produces a continual stream of fresh material onto orbits typical of both the Centaurs and the closely linked Scattered Disk objects. We note, however, that objects which initially move onto Scattered Disk orbits (with perihelion just beyond the orbit of Neptune) typically enter the Centaur region (perihelion within the orbit of Neptune) on quite short time scales. Therefore, it is fair to consider that the great majority of the population leaving the Trojan clouds will become Centaurs.

This result is not merely limited to theoretical clouds of Trojans dating back to the formation of the Solar system. We have also carried out the most detailed dynamical studies of the Neptune Trojan 2001 QR322 to date, in which 19683 test particles, spread throughout the observational 3σ error ellipse on the objects orbit, were followed under the influence of Jupiter, Saturn, Uranus and Neptune for a period of 1 Gyr. In contrast to the findings of earlier works (Marzari, Tricarico & Scholl 2003; Brasser et al. 2004) (a likely result of the ongoing refinement of the object's orbit), we discovered that the object appears to be somewhat dynamically unstable, with a decay half-life of ~550 Myr from the Trojan cloud, and ~590 Myr from the Solar system. Our results are presented in detail in Horner & Lykawka (2009), but allow us to here strengthen our conclusions on the eventual fate of objects once they leave the Neptunian Trojan cloud.

**3 RESULTS**
Figure 1 shows the rate at which the number of objects in the Neptunian Trojan region decays over the first 1 Gyr following the formation of the Trojan cloud, for four different initial formation scenarios (the results of detailed simulations of the formation and evolution of the Neptune Trojans, which will be presented in a future work). The different initial Trojan clouds which were used to create Figure 1 were based on the simulations detailed in Lykawka et al. (2009a), in which we examined the role that Neptune's migration played in shaping the Neptune Trojan population. The black line shows the decay of a population of objects captured from a trans-Neptunian disk as Neptune migrated from 18.1 AU to its current location over a period of 50 Myr, while the blue line shows the decay of a population of objects transported by the planet from pre-formed Trojan orbits under the same conditions. The green line shows the equivalent decay for a population captured from the trans-Neptunian disk during the 50 Myr migration of Neptune from 23.1 AU to its currently location, while the red line shows the decay of a population captured from the disk as a result of Neptune making the same migration in just 5 Myr.

It clear that a significant number of objects in the clouds experience dynamical instability during the very earliest days of their evolution, and are rapidly ejected from Neptune's 1:1 MMR. However, a gradual decay in the number of objects continues over the 1 Gyr of evolution simulated. This decay is indicative of the fact that Trojans on orbits which are initially "stable" slowly relax until they escape from the Trojan cloud onto Neptune-encountering orbits. Since the libration of objects around the leading and trailing Lagrange points of a planet[2] is only truly stable within the restricted three-body problem, it is clear that, within our Solar system (a far more complex dynamical system), any Trojan will be unstable over a sufficiently long time-scale. In the case of the Neptune Trojans, this is no doubt exacerbated by the proximity of the 2:1 MMR with the planet Uranus, which will, over time, help to destabilise even the most rigidly bound Neptune Trojans. Such destabilisation typically occurs as due to perturbations resulting from both the overlapping of the characteristic libration frequency of that MMR with that of the 1:1 MMR, or through resonant effects on the other frequencies associated with the orbital evolution of the Trojans (Chirikov 1979; Lecar, Franklin & Holman 2001). As these objects diffuse away from the Lagrange points, they will eventually escape from their Trojan orbits, and move out onto Neptune-encountering orbits of significantly lower stability. Once on such orbits (which often feature ongoing sequences of short-term captures in the various MMRs of the outer planets, as can be seen in Horner et al. 2004b), they are dynamically identical to any other Centaur moving in that region, and can go on to meet a variety of fates including evolution to become a short-period comet, ejection from the Solar system, or even direct impact onto a planet.

Such behaviour is illustrated in Figures 2-7, which show the evolution of six clones of the first Neptune Trojan to be discovered, 2001 QR322. These test particles started life within the error ellipse of that Trojan's best fit orbit, and were followed under the gravitational influence of the four giant planets until they were ejected from the Solar system or hit one of the giant planets. Each particle was initially highly stable – remaining on an almost identical orbit (with merely periodic fluctuations in eccentricity and inclinations as they librated around Neptune's L4 Lagrange point) for between 20 Myr and 3.34 Gyr, until eventually the grip of the 1:1 MMR on the object relaxed and it escaped into the Centaur region. The left side of the plot shows the evolution of the particles during their lifetime as a stable Neptune Trojan, while the right hand panels show the behaviour of

---
[2] Libration refers to the periodic movement of an object about a centre of equilibrium (or libration) during its resonant motion with a planet. For the 1:1 MMR, two centres of libration are particularly important due to their long term stability, the so-called L4 and L5 Lagrange points, displaced +60° and -60° from the planet on its orbit (Murray & Dermott 1999).

the object after leaving the 1:1 MMR. It is immediately clear that, after leaving Neptune's control, these particles all behave in a manner indistinguishable from that of Centaurs (compare, for example, with the objects discussed in Horner et al., 2004b). The objects were chosen at random from those which left the Neptunian Trojan cloud, while the simulations of the remaining clones of 2001 QR322 were still ongoing, as examples of objects leaving the Trojan clouds on a variety of different timescales, and their behaviour is representative of how such objects move upon leaving the resonance. One key feature in the evolution of such objects after leaving the Trojan cloud is the frequency with which they experience short-term capture into various resonances within the outer Solar system. In particular, in Figures 2-7, a number of such resonant trappings can be seen (primarily in the MMRs of Uranus and Neptune). Similar behaviour has also been reported for objects in the Scattered Disk, which are regularly passed between trans-Neptunian MMRs (for details, see e.g. Lykawka & Mukai 2007b)

So – it is clear that escaping Neptune Trojans will become Centaurs (indeed, so would objects escaping from the Jovian Trojan clouds). How large a contribution could this make to the overall Centaur population?

The answer to this question can be found from the combination of two key facts – first, the true population of the Neptune Trojan family, and second, the average dynamical lifetime of objects in that cloud. As discussed above, the first of the known Neptune Trojans to be studied in great depth (2001 QR322) appears to have a dynamical half-life of around 600 Myr according to comprehensive ongoing simulations (Horner & Lykawka 2009, (*in prep*)). In other words, a population of objects moving on an orbit similar to that of 2001 QR322 would halve in number every 600 Myr.

To be conservative, we first assume that the population of the Neptune Trojan family, at the current day, is the same as that of the main belt of asteroids (in reality, it is likely that the Neptune Trojans are far more numerous). It has been suggested that the population of objects in the 1km size range within the main belt is between 700,000 and $1.7 \times 10^6$ (Tedesco & Desert 2002). So, for our first guess, we will assume the Neptune Trojan population, on this scale, is $10^6$ objects. Again, to be cautious, we assume that 2001 QR322 is an unusually unstable object, and the Trojan family as a whole has a decay half-life just over three times longer – i.e. 2 Gyr. In other words, over the last 2 Gyr, $10^6$ Neptune Trojans have gone on to become Centaurs (since half the objects would decay over a period of 2 Gyr, it stands to reason that the population 2 Gyr in the past was double the current one), a rate of 1 object every ~2000 year. For comparison, Horner et al. (2004a) suggested that the inward flux from the trans-Neptunian region to the Centaur population would have to be 1 object every 125 years to support the current short-period comet population (based on earlier estimates of the mean lifetime of short-period comets, and the number currently known, by Fernandez (1985) and Levison & Duncan (1994)). Therefore, with these particularly conservative constraints on the lifetime and total population of Neptune Trojans, they are still capable of supplying around 6% of the Centaur population.

However, when recent results on the size distribution of the Neptune Trojans (Chiang & Lithwick 2005; Sheppard & Trujillo 2006 and references therein) are taken into account, it is clear that the true number of Neptune Trojans may be significantly higher than that used above. If we, instead, consider a current population of $10^7$ kilometre-sized objects in the Trojan region (which may still be somewhat conservative), then, even with a conservative mean half-life of 2 Gyr for the decay of these objects, the ongoing flux to the Centaur region would become 1 object every ~200 years. Again, comparing this to the flux of one new Centaur every 125 years, this implies that the Neptune Trojans could actually be the dominant source of the Centaurs, with a contribution of order 60% the total flux.

Finally, let us consider a still less conservative scenario. If we assume that 2001 QR322 is a typical Neptune Trojan (in other words, that the decay half-life of the family is of order 600 Myr), and that the estimates of the Trojan population being an order of magnitude higher than the asteroid belt are true (an assumption in broad agreement with estimates of the distribution of Neptune Trojans as a function of their size (Sheppard & Trujillo 2006)). This, then, would imply that the Trojan family had sourced $10^7$ new Centaurs in the last 600 Myr – a rate of one new Centaur every 60 years! Such a flux reaffirms that the Neptune Trojans could easily represent the dominant source of the Centaur population, and may even be an indication that either 2001 QR322 is atypically unstable for a Neptune Trojan, or that the true Trojan population may be significantly smaller than otherwise thought! This illustrates the great need for future observations to increase the known population of the Neptunian Trojan cloud.

While such analysis and estimates are by their very nature "back of an envelope" guesstimates, they do highlight the fact that the unstable Neptune Trojans could be a significant source of new Centaurs, and hence that many of the short period comets we observe today originate in the Neptunian 1:1 mean motion resonance. There is also one piece of interesting observational evidence which may support this claim. It is well known that the known Centaurs display a wide range of colours, with some having among the reddest colours in the Solar system (e.g. Pholus), while others (such as Chiron) are far bluer in colour (see e.g. Cruikshank et al. 1998; Doressoundiram et al. 2005). Indeed, as can be clearly seen in figure 2 of Tegler et al. (2008), the distribution of colours among known Centaurs appears to be strongly bi-modal, with a clear separation between the blue and red members. In figure 5 of that work, the authors show, for comparison, the colours of a variety of objects strewn throughout the Solar system, noting that the Centaurs are the only objects interior to the Edgeworth-Kuiper belt that display a population of extremely red objects. Peixinho et al. (2003) discuss this bi-modality in more depth. They conclude that the apparent excess of blue Centaurs over that which would be expected were resurfacing the sole mechanism making objects bluer requires a continual injection of fresh "blue" Centaurs from the Edgeworth-Kuiper belt. However, the great majority of objects in the classical EK-belt are strongly reddened (figure 5, in Tegler et al. 2008), and while the Plutinos (objects trapped in the 3:2 MMR with Neptune) and Scattered Disk objects do contain more blue objects, the distributions seem incompatible with that of the known Centaurs. However, observations of four of the known Neptune Trojans (and also, observations of the Jovian Trojans) reveal all examined objects to be blue (Sheppard & Trujillo 2006). If the Trojans are a significant source of Centaur material, then, we would expect them to significantly enhance the population of blue Centaurs compared with those that are red, which appears to be wholly compatible with the observed objects. This result fits nicely with the statistical result in Table 2 of Sheppard & Trujillo's 2006 work, which suggests that the objects with closest fit to the observational colours of the four observed Trojans were the blue Centaurs. Again, although it would be nice to have significantly more observational data to draw upon, it seems that there is observational evidence supporting a flux of objects from the Neptunian and Jovian Trojan clouds to the Centaur region.

**4 CONCLUSIONS**
We show that simulations of both hypothetical Neptune Trojans, formed both in that planet's Trojan clouds, and captured as it migrates, and simulations of the first known Neptune Trojan (2001 QR322) lead to the conclusion that members of the Neptune Trojan family regularly escape to help maintain the dynamically unstable Centaur population, the source of the great majority of short period comets. It is likely that similar decay processes operate on the Jovian Trojan clouds, sourcing material to the inner Solar system regularly over the age of the Solar system, as was suggested by Levison, Shoemaker & Shoemaker (1997) and Jewitt, Sheppard & Porco (2004). Observational evidence (particularly the surprisingly blue colours observed for objects in these two Trojan clouds, and the strong, unexplained blue component of the Centaur population) seems to support this conclusion. Future observational programs will prove vital in determining the true scale of the

Trojan contribution to the Centaur population. All sky survey programme (such as Pan-STARRS (Jewitt 2003)) will greatly improve the catalogue of known Centaurs and Trojans, while observations with platforms such as *HERSCHEL* (Mueller et al. 2009), which will obtain information on the physical and chemical properties of objects throughout the outer Solar system, will prove a key test of these theories.

**ACKNOWLEDGEMENTS**
PSL and JAH gratefully acknowledge financial support awarded by the Daiwa Anglo-Japanese Foundation and the Sasakawa Foundation, which proved vital in arranging an extended research visit by JAH to Kobe University. PSL appreciates the support of the COE program and the JSPS Fellowship, while JAH appreciates the ongoing support of STFC.

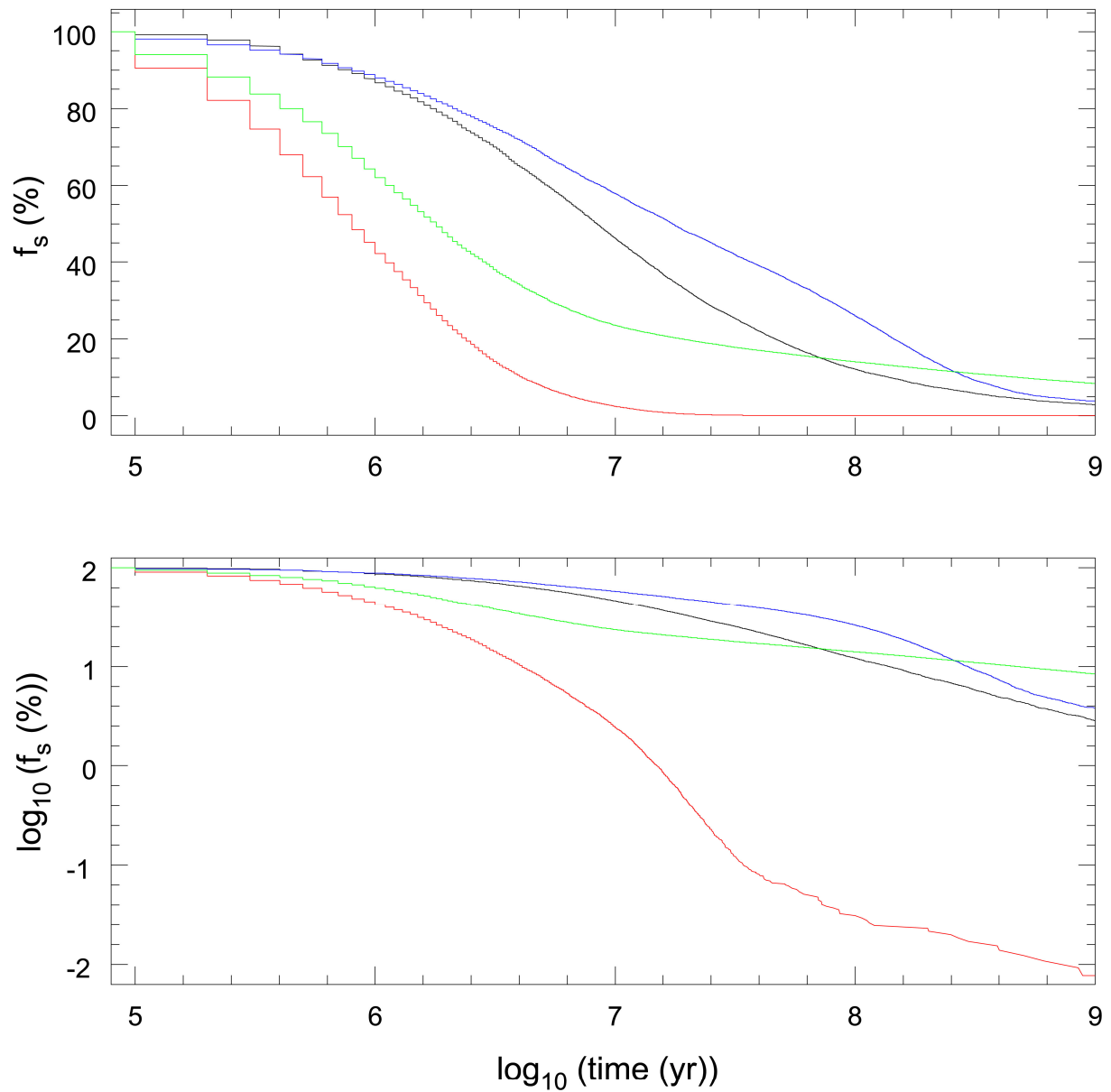

**Figure 1**: The decay of hypothetical Neptune Trojans as a function of time. The four lines represent Trojan populations created as a result of Neptune's migration during the latter stages of planet formation. The families produced were then followed for a period of 1 Gyr using *MERCURY* orbital integrator (Chambers 1999). Detailed results from this work will be published at a later date. The upper plot shows the percentage of Trojans remaining as a function of time (plotted logarithmically), while the lower plot shows the relationship between the surviving percentage and time when both are plotted logarithmically. What is clear is that, in each case, objects continually "leak" from the Neptune Trojan population, escaping into the domain of the Centaurs, a process we contend continues to the current day.

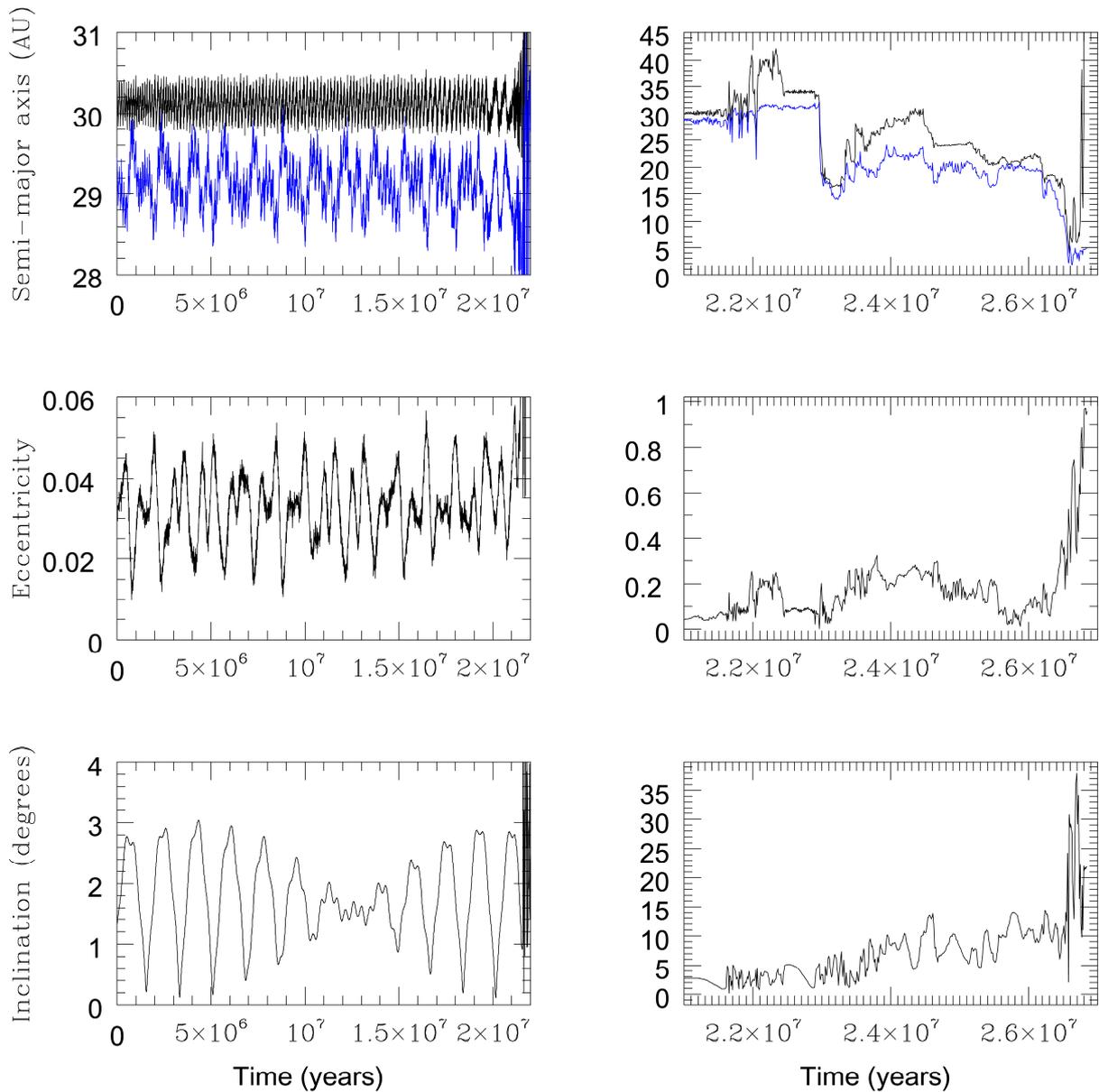

**Figure 2:** The dynamical behaviour of a clone of 2001 QR322, which spends just over 20 Myr as a Trojan before escaping from the L4 Trojan cloud. After its escape, it hops through a number of short resonant captures (such as that between 22.5 and 23 Myr) as it evolves through the Centaur region, finally becoming a Jupiter-family comet before its ejection from the Solar system by Jupiter. The left hand panels show the evolution of the semi-major axis (black, upper panel), perihelion distance (blue, upper panel), eccentricity (centre panel) and inclination (lower panel) over the 22 Myr of stable Trojan behaviour, while the right hand panels show the same variables after the object leaves the Trojan cloud. The Centaur behaviour of the object is clearly seen in these panels.

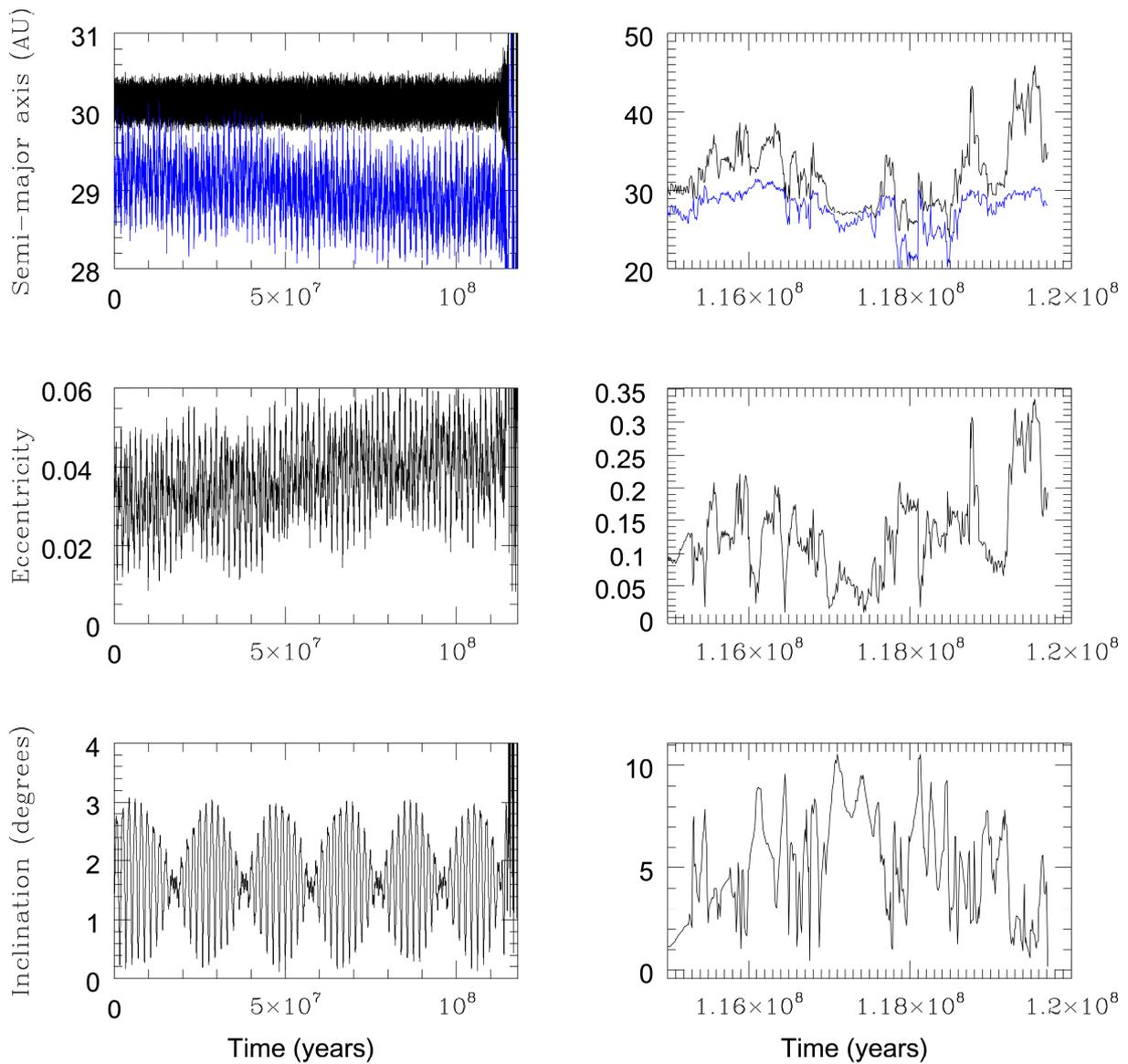

**Figure 3:** Dynamical behaviour of a clone of 2001 QR322, which remains a Neptune Trojan for just under 115 Myr before escaping, spending 5 Myr as a Centaur, and finally collides with Neptune. The left hand panels show the evolution of the semi-major axis (black, upper panel), perihelion distance (blue, upper panel), eccentricity (centre panel) and inclination (lower panel) over the 115 Myr of stable Trojan behaviour, while the right hand panels show the same variables after the object leaves the Trojan cloud. The Centaur behaviour of the object is clearly seen in these panels.

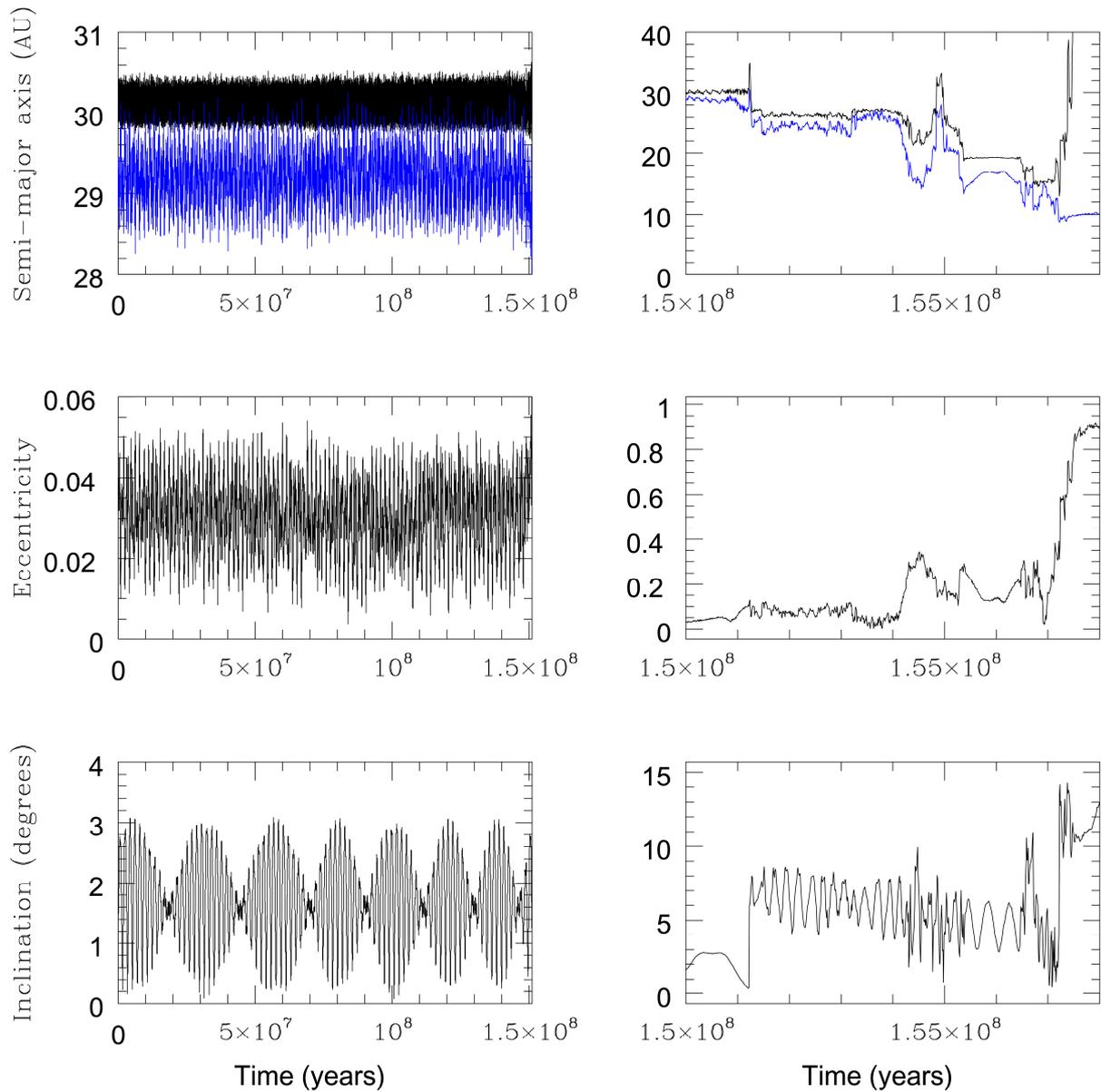

**Figure 4:** Dynamical behaviour of a clone of 2001 QR322 which remains a Neptune Trojan for approximately 151 Myr before escaping and spending a period of 7 Myr as a Centaur (including a temporary capture as a Uranus Trojan, just after 155 Myr). The object is eventually ejected from the Solar system by a series of close encounters with the planet Saturn. The left hand panels show the evolution of the semi-major axis (black, upper panel), perihelion distance (blue, upper panel), eccentricity (centre panel) and inclination (lower panel) over the 155 Myr of stable Trojan behaviour, while the right hand panels show the same variables after the object leaves the Trojan cloud. The Centaur behaviour of the object is clearly seen in these panels.

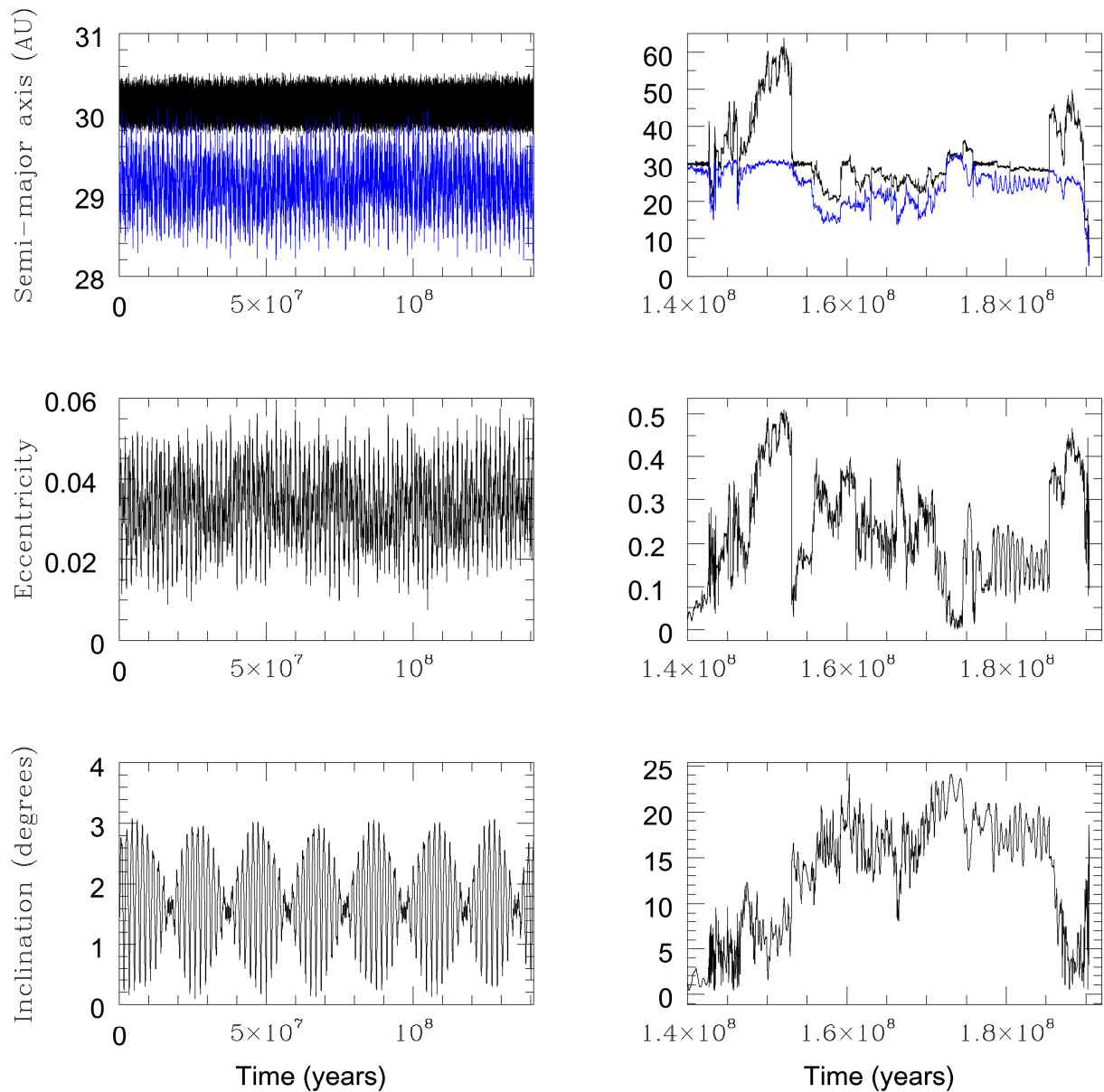

**Figure 5:** The dynamical behaviour of a clone of 2001 QR322, which spends a period of just over 140 Myr as a Neptune Trojan before escaping, and spending the next 50 Myr as a Centaur, until it briefly experiences capture as a Jupiter Family Comet before colliding with the planet Jupiter. The left hand panels show the evolution of the semi-major axis (black, upper panel), perihelion distance (blue, upper panel), eccentricity (centre panel) and inclination (lower panel) over the 140 Myr of stable Trojan behaviour, while the right hand panels show the same variables after the object leaves the Trojan cloud. The Centaur behaviour of the object is clearly seen in these panels.

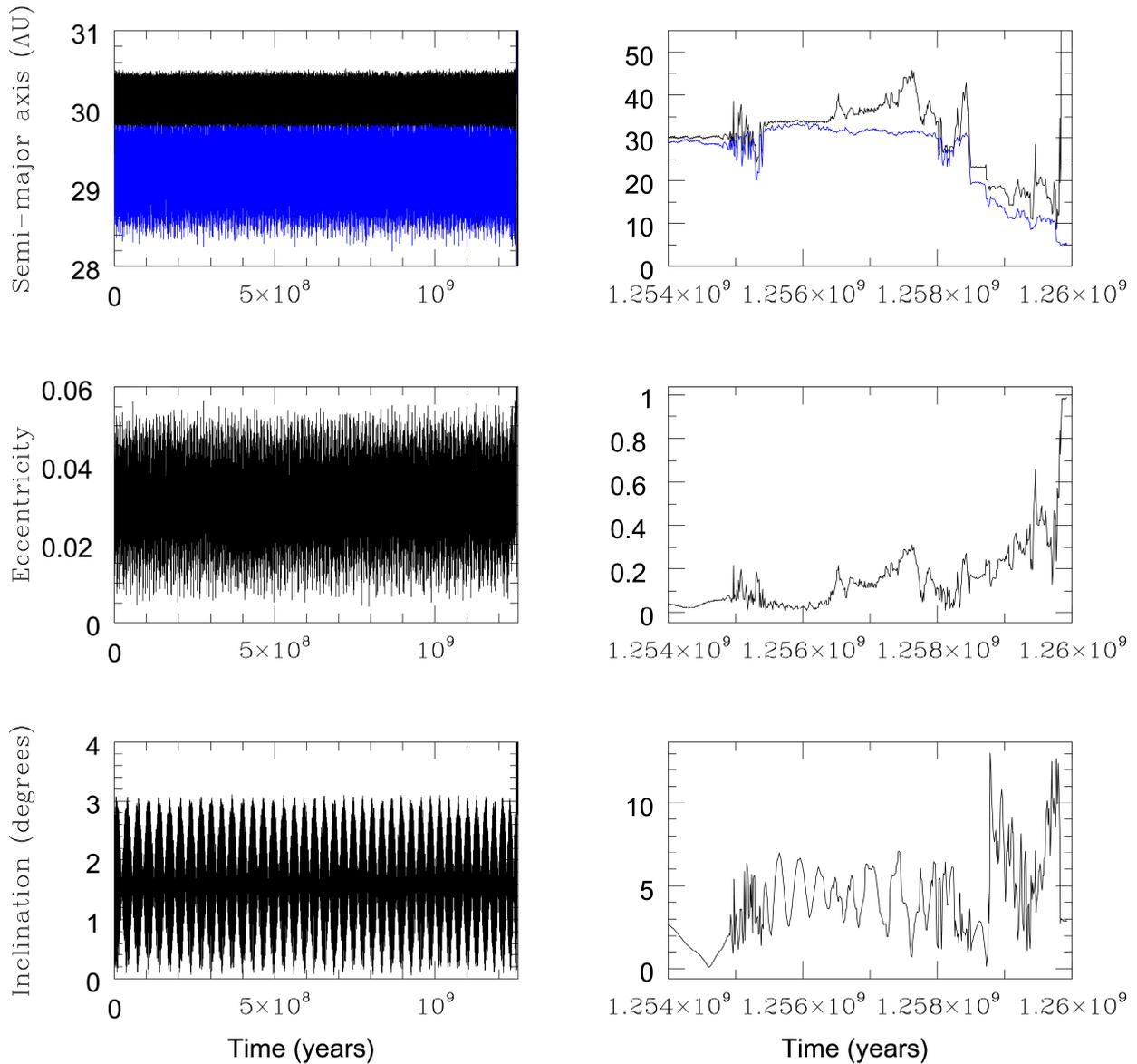

**Figure 6:** The dynamical behaviour of a clone of 2001 QR322, which spends 1.25 Gyr as a Trojan before escaping from the L4 Trojan cloud. After its escape, it undergoes a period of chaotic evolution at low eccentricity before being captured in a trans-Neptunian mean motion resonance. After leaving that resonance, its evolution continues to be dominated by Neptune until if is eventually handed down, and into a Uranian MMR. Finally, the object works its way down to Jupiter, which rapidly ejects it from the Solar system. The left hand panels show the evolution of the semi-major axis (black, upper panel), perihelion distance (blue, upper panel), eccentricity (centre panel) and inclination (lower panel) over the 1.25 Gyr Myr of stable Trojan behaviour, while the right hand panels show the same variables after the object leaves the Trojan cloud. The Centaur behaviour of the object is clearly seen in these panels.

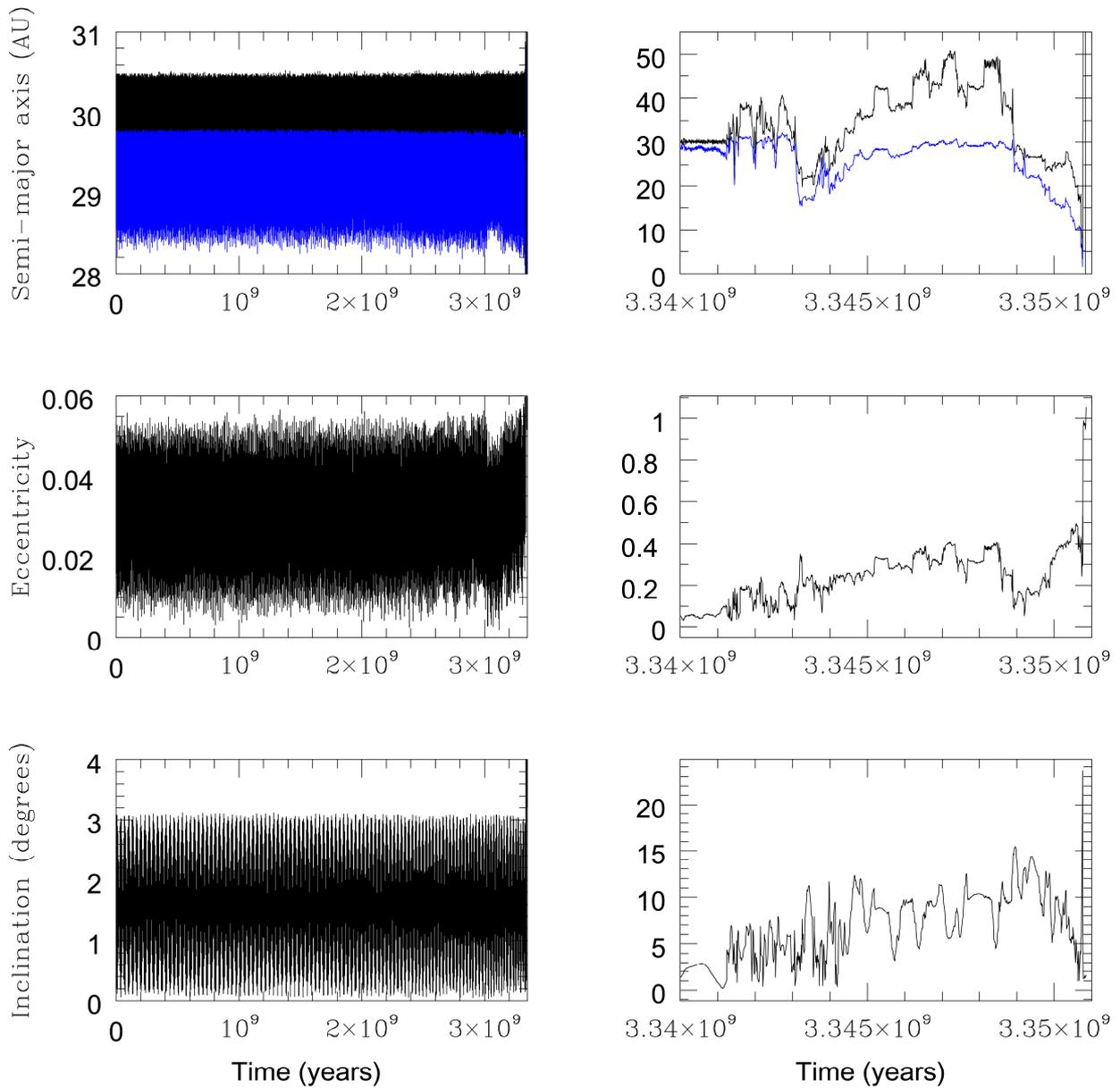

**Figure 7:** The dynamical behaviour of a clone of 2001 QR322, which spends in excess of 3.34 Gyr as a Trojan before escaping from the L4 Trojan cloud. After its escape, it spends almost 10 Myr as a chaotically evolving Centaur before eventually being thrown onto a short period cometary orbit. Finally, the object is ejected from the Solar system after a close encounter with Jupiter. The left hand panels show the evolution of the semi-major axis (black, upper panel), perihelion distance (blue, upper panel), eccentricity (centre panel) and inclination (lower panel) over the 3.34 Gyr of stable Trojan behaviour, while the right hand panels show the same variables after the object leaves the Trojan cloud. The Centaur behaviour of the object is clearly seen in these panels.